%Paper: astro-ph/9304024
%From: pradip@iopb.ernet.in
%Date: Tue Apr 27 01:45:49 1993

\input vanilla.sty
\scaletype{\magstep1}
\baselineskip 12pt

\define\nlb{\par\vskip 14truept}

\def\cl{\centerline}

\parindent = 20truept
\predisplaypenalty =0
\abovedisplayskip = 3mm plus 6truept minus 4truept
\belowdisplayskip = 3mm plus 6truept minus 4truept
\abovedisplayshortskip = 0mm plus 6truept
\belowdisplayshortskip = 2mm plus 6truept minus 4truept
\def\abstract#1 {\par{\narrower\noindent {\bf Abstract.} \hskip 2mm #1\par}}
\def\figure#1#2 {\par{\narrower\noindent {\bf Figure #1.} \hskip 2mm #2\par}}
\def\table#1#2 {\par{\narrower\noindent {\bf Table #1.} \hskip 2mm #2\par}}
\def\longref#1 {\par{\hangindent=20pt \hangafter=1 #1 \par}}

\def\begref{\vskip1mm\begingroup\let\INS=N}
\def\ref{\goodbreak\if N\INS\let\INS=Y\vbox{\noindent{\tenbf
}}\fi\hangindent\parindent
\hangafter=1\noindent\ignorespaces}
% End of References
\parskip 20pt
\tabskip=1em plus 2em minus .5em
\def\tablerule{\noalign{\hrule}}
\cl{\bf EIGENFREQUENCIES OF RADIAL PULSATIONS OF}
\cl{\bf STRANGE QUARK STARS}
\vskip 1.5 cm
\cl{B. Datta$^1$, Pradip K. Sahu$^2$, J. D. Anand$^3$ and A. Goyal$^3$}
\vskip 1cm
\cl{$^1$ Indian Institute of Astrophysics}
\cl{Bangalore 560034, India}
\vskip .5cm
\cl{$^2$ Institute of Physics}
\cl{Bhubaneswar 751005, India}
\vskip .5cm
\cl{$^3$ Department of Physics}
\cl{University of Delhi}
\cl{Delhi 110007, India}
\vskip .5 cm
\noindent {\cl {ABSTRACT}}

We calculate the range of eigenfrequencies of radial pulsations of stable
strange quark stars, using the general relativistic pulsation
equation and adopting realistic equation of state for degenerate
strange quark matter.
\vskip 1.5 cm
The equation governing infinitesimal radial pulsations of a
nonrotating star in general relativity was given by Chandrasekhar
[1], and it has the following form :

$$ F \frac {d^2\xi}{dr^2} + G \frac{d\xi}{dr} + H\xi = \sigma^2\xi ,
\eqno (1) $$

\noindent where $\xi$(r) is the Lagrangian fluid displacement and
c$\sigma$ is the characteristic eigenfrequency.  The quantities F, G,
H depend on the equilibrium profiles of the pressure$(p)$ and
density$(\rho)$  of
the star, and are given by

$$ F = - e^{-\lambda} e^{\nu} \Gamma p/(p+\rho c^2) \eqno (2) $$

$$\eqalign{G = - e^{-\lambda} e^{\nu} \Bigl\{&\Gamma p \Big(\frac {1}{2}
\frac{d\nu}{dr} + \frac{1}{2} \frac{d\lambda}{dr} + \frac{2}{r}\Big) + \cr
&p \frac{d\Gamma}{dr} + \Gamma \frac{dp}{dr}\Bigr\}/(p+\rho c^2)}
\eqno (3) $$

$$ H = \frac {e^{-\lambda}e^{\nu}}{p + \rho c^2} \Bigl\{ \frac{4}{r}
\frac{dp}{dr} - \frac{(dp/dr)^2}{p + \rho c^2} - A \Bigr\} + \frac
{8\pi G}{c^4} e^{\nu} p \eqno (4) $$

\noindent where $\Gamma$ is the adiabitic index, defined in the
general relativistic case as

$$ \Gamma = (1 + \rho c^2/p) \frac{dp}{d(\rho c^2)} \eqno (5) $$

\noindent and

$$\eqalign{A = &\frac {d\lambda}{dr} \frac{\Gamma p}{r} + \frac
{2p}{r} \frac{d\Gamma}{dr} + \frac {2\Gamma}{r} \frac{dp}{dr} - \frac{2\Gamma
p}{r^2}\cr
& - \frac{1}{4} \frac{d\nu}{dr} \Big( \frac{d\lambda}{dr} \Gamma p +
2p \frac{d\Gamma}{dr} + 2\Gamma \frac{dp}{dr} - \frac{8\Gamma p}{r}\Big)\cr
& - \frac{1}{2} \Gamma p \Big(\frac{d\nu}{dr}\Big)^2 - \frac{1}{2}
\Gamma p \frac{d^2\nu}{dr^2}} \eqno (6)$$

The boundary conditions to solve the pulsation equation (1) are

$$ \xi (r = 0) = 0 \eqno (7)$$
$$ \delta p (r = R) = -\xi \frac{dp}{dr} - \Gamma p \frac
{e^{\nu/2}}{r^2} \frac{\partial}{\partial r} (r^2 e ^{-\nu/2} \xi)\vert_r=R = 0
\eqno (8) $$

\noindent Eq. (1) is of the Sturm - Liouville type and has real eigenvalues

$$ \sigma^2_o < \sigma^2_1 < \dots < \sigma^2_n < \dots \dots , $$

\noindent with the corresponding eigenfunctions $\xi_o$(r),
$\xi_1$(r), \dots $\xi_n$(r), \dots , where
$\xi_n$(r) has n nodes.

At high baryonic densities, bulk strange matter is in an overall
colour singlet state, and can be treated as a relativistic fermi gas
interacting perturbatively, the quark confinement property being
simulated by the phenomenological bag model constant (B).  Chemical
equilibrium between the three quark flavours and electrical charge
neutrality allow us to calculate the EOS from the thermodynamic
potential of the system as a function of the quark masses, the bag
pressure term (B) and the renormalization point $\mu_o$. To second order
in $\alpha_c$, and assuming u and d quarks to be massless, the
thermodynamic potential is given by [2] :

$$ \Omega = \Omega_u + \Omega_d + \Omega_s + \Omega_{int.} +
\Omega_e, \eqno (9) $$

\noindent where $\Omega_i$ (i = u, d, s, e) represents the contributions of
u, d, s quarks and electrons  and
$\Omega_{int}$ is the contribution due to interference between u and d
quarks and is of order $\alpha^2_c$ :

$$ \Omega_u = - \frac{1}{4\pi^2} \mu^4_u \Bigl[ 1 - \frac{8 \alpha_c}
{\pi} - 16 \Big(\frac{\alpha_c}{\pi}\Big)^2 ln \Big(\frac{\alpha_c}
{\pi}\Big) - 31.3 \Big(\frac{\alpha_c}{\pi}\Big)^2 \Bigr] \eqno
(10) $$

$$ \Omega_d = \Omega_u \big(\mu_u \leftrightarrow \mu_d\big) \eqno
(11) $$

$$ \eqalign{ \Omega_s = - \frac{\mu^4_s}{4\pi^2} \Big\{& (1 -
\lambda^2)^{1/2} (1 - \frac{5}{2} \lambda^2) + \cr
& \frac{3}{2} \lambda^4 ln \big\{ [1 + (1 - \lambda^2)^{1/2}] /
\lambda\big\}\cr
& - \frac{8\alpha_c}{\pi} \big[ 3\Big( (1 - \lambda^2)^{1/2} - \lambda^2 ln
\big\{ [1 + (1 - \lambda^2)^{1/2}] / \lambda\big\}\Big)^2 \cr
& - 2 (1 - \lambda^2)^2 \big]\Big\}} \eqno (12)$$

$$\eqalign {\Omega_{int.} = &\frac{1}{\pi^2} \Big(\frac{\alpha_c}
{\pi}\Big)^2 \Big\{8 \mu^2_u \mu^2_d ln \Big(\frac{\alpha_c}{\pi}\Big) -
1.9 \mu^2_u \mu^2_d \cr
& - 19.3 \Big\{ \mu^4_u ln \big[ \mu^2_u / ( \mu^2_u + \mu^2_d)\big]
+ \cr
& \mu^4_d ln \big[ \mu^2_d / (\mu^2_u + \mu^2_d) \big]\Big\}\cr
& - 4 (\mu^2_u + \mu^2_d)\Big\{\mu^2_u ln \big[\mu^2_u / (\mu^2_u +
\mu^2_d)\big]\cr
& + \mu^2_d ln \big[\mu^2_d / (\mu^2_u + \mu^2_d)\big]\Big\}\cr
& + \frac {4}{3} \big(\mu_u - \mu_d\big)^4 ln \Big[ \vert\mu^2_u -
\mu^2_d \vert / (\mu_u \mu_d)\Big] \cr
& + \frac{16}{3} \mu_u \mu_d \big(\mu^2_u + \mu^2_d\big) ln
\Big[(\mu_u + \mu_d)^2 / (\mu_u \mu_d)\Big]\cr
& - \frac{4}{3} \big(\mu^4_u - \mu^4_d\big) ln
\big(\frac{\mu_u}{\mu_d}\big) \Big\}} \eqno (13)$$

$$ \Omega_e = - \frac{\mu_e^4}{12 \pi^2}. \eqno (14) $$

\noindent Here $\mu_i$ is the chemical potential of the ith particle
species and $\lambda = m_s / \mu_s$.  We neglect the strange-quark
contribution to order $\alpha^2_c$ and higher
in the thermodynamic potential $\Omega_s$.
The screened charge $\alpha_c$ is obtained by solving the
Gell-Mann$-$Low equation [2] :

$$ \mu \frac{d\alpha_c (\mu)}{d\mu} =  \Big\{ - \frac{58}{3\pi} -
8\pi\mu \frac{d}{d\mu}\pi_s(\mu) \Big\} \alpha^2_c -
\frac{460}{3\pi^2} \alpha^3_c(\mu) , \eqno (15)$$
\noindent which includes the effects of the strange-quark mass in the
lowest order.  The higher order contribution to the Gell-Mann$-$Low
equation due to strange quarks may be ignored because these
are important only at low densities where the
coupling is strong but the pair production of massive strange quarks
is unimportant (see ref. [2] for further discussions).

The vacuum polarization tensor $\pi_s(\mu)$ for the strange quarks  is
given by

$$\eqalign{\pi_s(\mu) = \frac{1}{4\pi^2} &\Big\{ \frac{5}{9} - \frac
{4m^2_s}{3\mu^2} - \frac{2}{3} \big[\big(1-2m^2_s\big) / \mu^2\big]
\times \cr
& \big(1 + 4m^2_s / \mu^2\big)^{1/2} \ \ \hbox{arctanh} \ \ \ \big(1
+ 4m^2_s / \mu^2\big)\Big\}^{-1/2}} \eqno (16)$$

\noindent In Eq. (15), $\alpha_c(\mu_o)$ is the value of $\alpha_c$ at
the renormalization point $\mu_o$, where it is taken to be equal to 1.

The total energy density and the external pressure of the system are
given by

$$\eqalign{\epsilon & = \Omega + B + \sum_i \mu_i n_i}\eqno (17) $$

$$\eqalign{p & = - \Omega - B} \eqno (18) $$

\noindent where $n_i$ is the number density of the ith particle
species.  For specific choices of the parameters of the theory
(namely, m$_s$, B and $\mu_o$), the EOS is now obtained by calculating
$\epsilon$ and p for a given value of $\mu$ :

$$ \mu \equiv \mu_d = \mu_s = \mu_u + \mu_e \eqno (19) $$

\noindent by solving for $\mu_e$ from the condition that the total
electric charge of the system is zero.

There is an unphysical dependence of the EOS on the renormalization
point $\mu_o$, which, in principle, should not affect the
calculations of physical observables if the calculations are
performed to all orders in $\alpha_c$ [3,4].  In practice, the calculations
are done perturbatively and, therefore, in order to minimize the
dependence on $\mu_o$ the renormalization point should be chosen to
be close to the natural energy scale, which could be either $\mu_o
\simeq B^{1/4}$ or the average kinetic energy of quarks in the bag, in
which case, $\mu_o \simeq$ 313 MeV. In the present study, our choise
of $\mu_o$ is dictated by the requirement that stable strange matter
obtains at zero temperature and pressure with a positive baryon
electric charge [5].  This leads to the following representative choice
of the parameter values : EOS model 1 : B = 56 MeV fm$^{-3}$; m$_s$ = 150 MeV;
$\mu_o$ = 150 MeV. EOS model 2 : B = 67 MeV fm$^{-3}$' m$_s$ = 150 MeV;
$\alpha_c$ = 0. Model 2 corresponds to no quark interactions, but a non-zero
mass
for the strange quark.

Numerical values of pressure (p) and total mass-energy density ($\rho
= \varepsilon/c^2$) for the quark matter EOS models used here are
listed in Table 1.  For the sake of comparison, we have included in
this table the EOS corresponding to non-interacting, massless quarks
as given by the simple MIT bag model with B = 56 MeV fm$^{-3}$.
Among these EOS, the bag model is `stiffest' followed by models 1 and
2. Equilibrium configurations of strange quark stars,
corresponding to the above EOS, are
presented in Table 2, which lists the gravitational mass (M), radius
(R), obtained  by integrating the relativistic stellar structure
equations the surface redshift (z) given by

$$ z = (1 - 2 GM/c^2R)^{-1/2} - 1 \eqno (20) $$

\noindent and the period (P$_o$) corresponding to the fundamental
frequency $\Omega_o$ defined as [6] :

$$\Omega_o = \big(3 GM/4R^3\big)^{1/2} \eqno (21)$$

\noindent as functions of the central density ($\rho_c$) of the star.

We solved Eq. (1) for the eigenvalue $\sigma$ by writing the
differential equation as a set of difference equations.  The
equations were cast in tridiagonal form and the eigenvalue found by
using the EISPACK routine.  This routine finds the eigenvalues of a
symmetric tridiagonal matrix by the implicit QL method.

Results for the oscillations of quark stars corresponding to EOS
models 1 and 2 are illustrated in Fig. 1.  For purpose of comparison,
we have included in Fig.1 the results for quark stars corresponding
to (a) the simple MIT bag EOS (non-interacting, massless quarks and B
= 56 MeV fm$^{-3}$) and also (b) neutron stars corresponding to a
recently given neutron matter EOS [7].  The plots in Fig.1 are for
the oscillation time period (= 2$\pi/c\sigma$) versus the
gravitational mass (M).  The fundamental mode and the first four
harmonics are considered.  The period is an increasing function of M,
the rate of increase being progressively less for higher oscillation
modes.  The fundamental mode  oscillation periods for quark stars are
found to have the following range of values :
\settabs \+ MIT bag model \quad & :\quad & (0.14 - 0.32) milliseconds
\quad \cr
\+ MIT bag model & : & (0.14 - 0.32) milliseconds \cr
\+ EOS model 1 & : & (0.10 - 0.27) \quad ''\cr
\+ EOS model 2 & : & (0.06 - 0.30) \quad ''\cr
\noindent For neutron stars, we find that the range of
periods for the l = 0 mode is  $\simeq$ 0.3 milliseconds.
For the higher modes, the periods are $\leq$ 0.2 milliseconds,
similar to the case of quark stars.

Inclusion of strange quark mass and the quark interactions make the
EOS a little `softer' as compared to the simple MIT bag EOS (see
Table 1).  This is
reflected in the value of the maximum mass of the strange quark star
(see Table 2).
For the pulsation of quark stars this gives, for l=0 mode
eigenfrequencies, values as low as 0.06 milliseconds.
The main conclusion that emerges from our study,
therefore, is that use of realistic EOS can be important in deciding
the range of eigenfrequencies, at least for the fundamental mode of
radial pulsation.  The
results presented here thus form an improved first step of calculations on
radial oscillations of neutron stars with a quark matter core
presented by Haensel et al. [8], whose numerical
conclusions are expected to get altered.
\vfill\eject
\noindent {\bf References}
\item{1.} S.Chandrasekhar, Astrophys. J. {\bf 140} (1964) 417.
\item{2.} B.Freedman and L.McLerran, Phys. Rev. D {\bf 17} (1978) 1109.
\item{3.} C.Alcock, E.Farhi and A.Olinto, Astrophys. J. {\bf 310}
(1986) 261.
\item{4.} A.Goyal and J.D.Anand, Phys. Rev. D {\bf 42} (1990) 992.
\item{5.} V.Baluni, Phys. Lett. B {\bf 72} (1978) 381; Phys. Rev. D
{\bf 17} (1978) 2092.
\item{6.} C.Cutler, L.Lindblom, and R.J.Splinter, Astrophys. J.
{\bf 363} (1990) 603.
\item{7.} R.B.Wiringa, V.Fiks, and A.Fabrocini, Phys. Rev. C {\bf
38} (1988) 1010.
\item{8.} P.Haensel and A.J.Jerzak, Acta Phys. Polonica B {\bf 20}
(1989) 141.
\vfill\eject
\cl {\bf Table 1}
\nlb
\cl{EQUATIONS OF STATE FOR DEGENERATE STRANGE QUARK MATTER}
\vskip 1cm
\hrule
\halign to \hsize{\hfil#\hfil &\hfil#\hfil & \hfil#\hfil & \hfil#\hfil\cr
&&&\cr
$\rho$ &\omit\multispan3 \hfil{P(10$^{36}$ dynes cm$^{-2}$)}\hfil\cr
(10$^{14}$ g cm$^{-3})$ & model 1 & model 2 & MIT Bag (B=56 MeV fm$^{-3}$)\cr
&&&\cr
\tablerule
&&&\cr
6.0  & 4.44 & 2.23 & 6.01 \cr
8.0 & 10.13 & 7.83 & 12.00 \cr
10.0 & 15.88 & 13.49 & 17.99 \cr
12.0 & 21.63 & 19.17 & 23.99 \cr
14.0 & 27.41 & 24.88 & 29.98 \cr
&&&\cr
16.0 & 33.20 & 30.16 & 35.97 \cr
18.0 & 39.00 & 36.36 & 41.96 \cr
20.0 & 44.82 & 42.12 & 47.95 \cr
22.0 & 50.64 & 47.89 & 53.95 \cr
24.0 & 56.47 & 53.67 & 59.94 \cr
&&&\cr
26.0 & 62.30 & 59.46 & 65.93 \cr
28.0 & 68.14 & 65.26 & 71.92 \cr
30.0 & 73.98 & 71.06 & 77.91 \cr
32.0 & 79.83 & 76.87 & 83.90 \cr
36.0 & 91.53 & 88.51 & 95.89 \cr
&&&\cr
40.0 & 100.32 & 100.16 & 107.87 \cr
50.0 & 132.58 & 129.36 & 137.83 \cr
&&&\cr}\hrule
\vfill\eject
\cl{\bf Table 2}
\nlb
\cl{EQUILIBRIUM STRANGE QUARK STAR MODELS}
\vskip 1cm
\hrule
\halign to \hsize{\hfil#\hfil & \hfil#\hfil & \hfil#\hfil
        &\hfil#\hfil & \hfil#\hfil & \hfil#\hfil \cr
&&&&&\cr
Equation & $\rho_c$ & M/M$_{\odot}$ & R & Surface & P$_o$\cr
of State &&&&redshift&\cr
& (10$^{14}$ g cm$^{-3}$) &&(km) & (z) &(milliseconds)\cr
&&&&&\cr
\tablerule
&&&&&\cr
Model 1 & 24.0 & 1.958 & 10.55 & 0.487 & 0.488 \cr
&20.0 & 1.967 & 10.78 & 0.472 & 0.503 \cr
&16.0 & 1.951 & 11.02 & 0.448 & 0.522 \cr
&12.0 & 1.864 & 11.22 & 0.401 & 0.548 \cr
&8.0 & 1.521 & 11.02 & 0.299 & 0.591 \cr
&6.0 & 0.997& 9.93  & 0.192 & 0.624 \cr
&5.0 & 0.485 & 7.99 & 0.104 & 0.646 \cr
&&&&&\cr
Model 2 & 24.0 & 1.863 & 10.09 & 0.483 & 0.468 \cr
&20.0 & 1.862 & 10.29 & 0.465 & 0.482 \cr
&16.0 & 1.829 & 10.49 & 0.435 & 0.500 \cr
&12.0 & 1.710 & 10.62 & 0.381 & 0.527 \cr
&8.0 & 1.281 & 10.14 & 0.263 & 0.568 \cr
&6.0 & 0.645 & 8.37 & 0.138 & 0.600 \cr
&5.0 & 0.092 & 4.48 & 0.032 & 0.622 \cr
&&&&&\cr
MIT Bag & 24.0 & 2.021 & 10.81 & 0.493 & 0.500 \cr
model &20.0& 2.033 & 11.04 & 0.480 & 0.514 \cr
(B=56 &16.0 & 2.023 & 11.29 & 0.450 & 0.533 \cr
MeV fm$^{-3}$)&12.0 & 1.947 & 11.52 & 0.410 & 0.558 \cr
&8.0 & 1.635 & 11.41 & 0.310 & 0.604 \cr
&6.0 & 1.150& 10.52 & 0.210 & 0.636 \cr
&5.0 & 0.666 & 8.98 & 0.130 & 0.659 \cr
&&&&&\cr}
\hrule
\vfill\eject
\cl{\bf Figure captions}
\vskip 1cm
\item{Figure 1.} Periods of radial pulsations as functions of the
gravitational mass.  The top two and bottom left boxes correspond to strange
quark
stars.  The bottom right box is for stable neutron stars
corresponding to beta-stable neutron matter, model UV14 + UVII, ref.
[7].  The labels 1, 2, 3, 4, 5 correspond respectively to the
fundamental and the first four harmonics.
\vfill\eject
\end